\documentstyle[12pt]{article}

 1
 1
 1

\def\ie{{\em i.e.}}
\def\eg{{\em e.g.}}

\def\beq{\begin{equation}}
\def\eeq{\end{equation}}
\def\bdm{\begin{displaymath}}
\def\edm{\end{displaymath}}

\catcode`\@=11 % This allows us to modify PLAIN macros.
\def\coeff#1#2{{\textstyle{#1\over #2}}}

\def\lsim{\mathrel{\mathpalette\@versim<}}
\def\gsim{\mathrel{\mathpalette\@versim>}}
\def\@versim#1#2{\vcenter{\offinterlineskip
    \ialign{$\m@th#1\hfil##\hfil$\crcr#2\crcr\sim\crcr } }}
\def\etal{{\em et. al.}}
\def\JL{J. L. Lopez}
\def\DVN{D. V. Nanopoulos}
\def\AZ{A. Zichichi}

\def\t1{{\tilde 1}}

\def\GeV{\,{\rm GeV}}
\def\TeV{\,{\rm TeV}}

\def\to{\rightarrow}

\def\NPB#1#2#3{Nucl. Phys. B {\bf#1} (19#2) #3}
\def\PLB#1#2#3{Phys. Lett. B {\bf#1} (19#2) #3}

\def\PRD#1#2#3{Phys. Rev. D {\bf#1} (19#2) #3}
\def\PRL#1#2#3{Phys. Rev. Lett. {\bf#1} (19#2) #3}
\def\PRT#1#2#3{Phys. Rep. {\bf#1} (19#2) #3}

\begin{document}
\begin{flushright}
CTP-TAMU-39/94\\
hep-ph/9408279\\
August 1994
\end{flushright}
\begin{center}
{\huge\bf Theoretical expectations for\\ the top-quark mass}
\end{center}
\vspace{0.25cm}
\begin{center}
{ JORGE L. LOPEZ\\}
\vglue 0.25cm
{\em Department of Physics, Texas A\&M University\\}
{\em College Station, TX 77843--4242, USA\\}
\end{center}
\vspace{0.25cm}
\begin{abstract}
I review the theoretical expectactions for the top-quark mass in a
variety of models: the Standard Model, unified models (GUTs), low-energy
supersymmetric models (SUSY), unified supersymmetric models (SUSY GUTs),
supergravity models, and superstring models. In all instances I consider the
constraints on the top-quark mass which arise by demanding that these theories
be weakly interacting. This assumption is quantified by the use of partial-wave
unitarity or triviality.  The resulting upper bounds on the top-quark mass are
most stringent in SUSY GUTs models ($m^{\rm pole}_t\lsim200\sin\beta\GeV$). I
also discuss a class of $SU(5)\times U(1)$ superstring models where $m^{\rm
pole}_t\sim(170-195)\GeV$ is predicted. I conclude that experimental
determinations of the top-quark mass can be {\em naturally} understood in SUSY
GUTs and superstring models. (Lecture presented at the International School of
Subnuclear Physics, 32nd Course: From Superstring to Present-Day Physics,
Erice--Sicily: July 3--11, 1994.)
\end{abstract}

\baselineskip=14pt

\section{Introduction}
On April 26, 1994 the CDF Collaboration announced ``evidence" for the existence
of the top quark in $p\bar p$ collisions at $\sqrt{s}=1.8\TeV$ \cite{CDF}.
In a lengthy paper CDF argued that they had observed an excess of dilepton
and lepton+jets events which were most naturally explained as coming from
$t\bar t$ production. A kinematical fit to the candidate event masses yields
$m_t=174\pm17\GeV$, when systematic and statistical errors are combined in
quadrature. In reality, the issue of the existence of the top quark is not
settled since the statistical significance of the measurement is not
compelling, there are annoying discrepancies among various observables
(\eg, the theoretically predicted QCD cross section for the determined
top-quark mass and the actually measured cross section), and the other detector
at the Tevatron (D0) does not observe any meaningful excess of top-like events.
Fortunately the Tevatron is currently running (Run IB) and should accumulate
five times as much data by the end of the run. This increased data sample
should shed a lot of light on the present top-quark controversy.

Indirect evidence for the existence of the top quark has been mounting through
fits to the LEP electroweak observables \cite{LEPmt} which depend on the value
of $m_t$. These fits can be compared with the theoretical calculations
\cite{EFLmt} and a value of $m_t$ is deduced within the framework of the
Standard Model. This value has a Higgs-boson mass uncertainty which is now
comparable to the experimental uncertainty. For light Higgs-boson masses
(\ie, those consistent with a supersymmetric Standard Model) the latest
fit gives $m_t=162\pm9\GeV$ \cite{EFL} (this value includes the CDF
measurement). Heavier Higgs-boson masses (\eg, $m_H=300\GeV$ as is many
times assumed) increase the central value by as much as 20 GeV. In any
event, either directly (CDF) or indirectly (LEP) there seems to be evidence
for the existence of the top quark. Moreover, its presumed mass makes it
the heaviest elementary particle ever discovered. In fact, we have
$m_t={\cal O}(M_{W,Z})$, which hints that the top quark may have something to
do with the breaking of the electroweak symmetry (\eg, as in the radiative
electroweak breaking mechanism in a supergravity theory).

\begin{figure}[p]
\vspace{6.7in}
\includegraphics{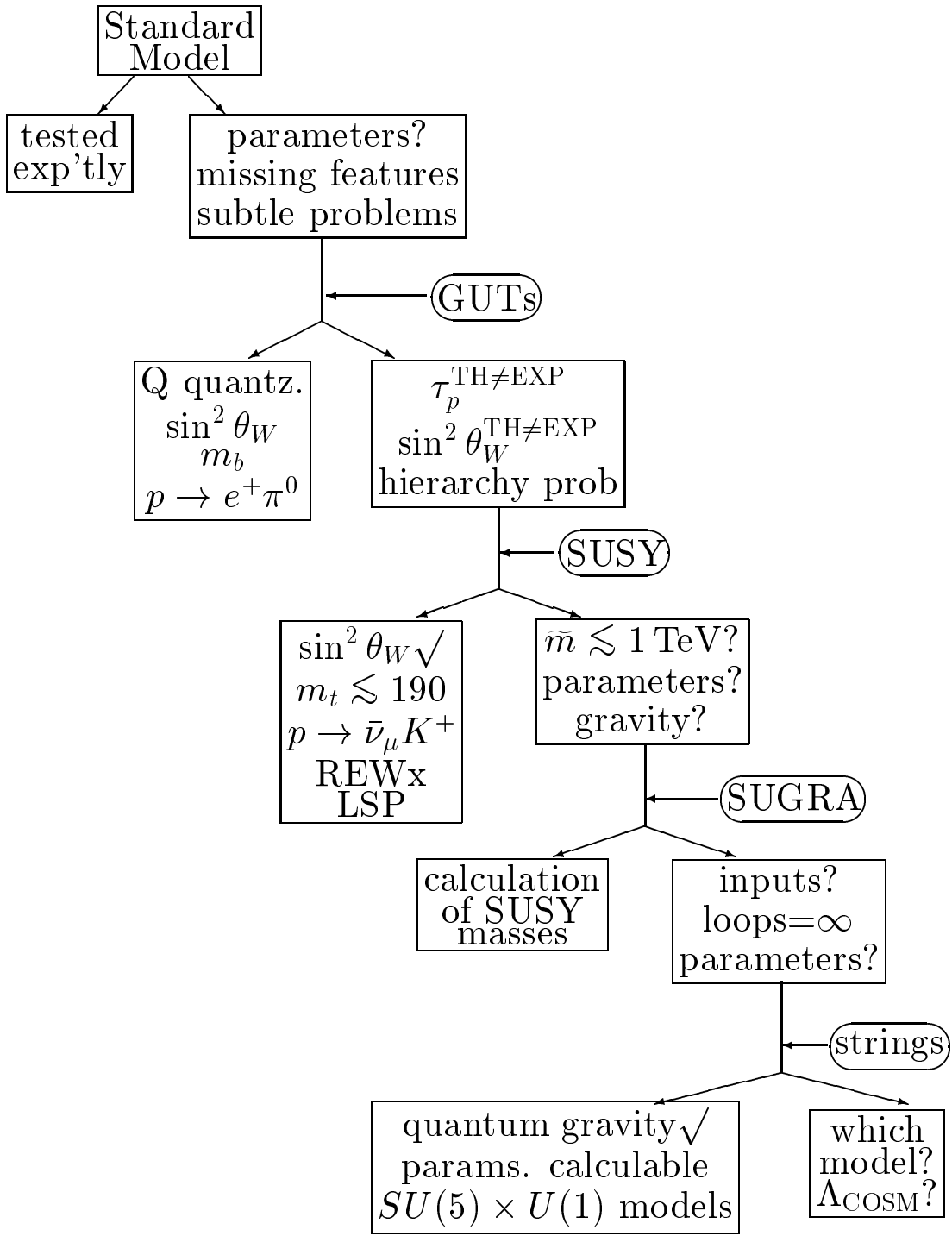}
\vspace{2cm}
\caption{A ``road map" showing the successes (left arrows) and problems
(right arrows) of a sequence of theoretical frameworks from the Standard
Model up to superstring models.}
\label{RoadMap}
\end{figure}

In what follows I will review the theoretical expectations for the top-quark
mass in a variety of theoretical frameworks: the Standard Model, unified models
(GUTs), low-energy supersymmetric models (SUSY), supersymmetric unified models
(SUSY GUTs), supergravity, and superstrings. The motivation for considering
such a sequence of theoretical frameworks is well known \cite{reviews} and
will not be elaborated any further. However, to refresh your memory in
Fig.~\ref{RoadMap} I show a ``road map" starting from the Standard Model and
ending at very high energies with superstring models.

The theoretical expectations which I consider are mostly in the form of upper
bounds on the top-quark mass. These follow from the practical and usually tacit
assumption that the various theories be weakly interacting in their regime of
applicability. Violation of these bounds entails a strongly interacting theory
which cannot be analyzed in the usual way. That is, we have a new ``phase" of
the theory with new physical consequences -- a different theory. In the case of
superstring models, one is able to calculate in principle the top-quark mass
and the expectations we discuss are not just upper bounds, but actual
predictions.

\section{The Standard Model}

For our present purposes I will view the Standard Model as an effective
theory valid for energies $\lsim1\TeV$. The top-quark mass is related
to the top-quark Yukawa coupling in the Yukawa Lagrangian via
\begin{equation}
m_t=\lambda_t {v_0\over\sqrt{2}}=174\lambda_t\GeV,
\label{mtSM}
\end{equation}
where $v_0=1/\sqrt{\sqrt{2}G_F}=246\GeV$ is the Higgs vacuum expectation value.
This relation shows that large values of $m_t$ originate from large values of
$\lambda_t$. However, the interactions among top quarks, or top quarks with
Higgs bosons and the longitudinal components of the electroweak gauge bosons
(\ie, the would-be Goldstone bosons) are all proportional to $\lambda_t$.
Therefore, a large
value of $m_t\leftrightarrow\lambda_t$ leads to strong interactions in the
Yukawa sector of the Standard Model. This new ``phase" of the Standard Model
has been investigated in the context of $t\bar t$ condensates \cite{Bardeen},
and may be fine in its own right,\footnote{Until the end of the lecture we will
not worry about whether the top-quark masses we are considering may already
be in conflict with experimental observations. We would like to focus on the
theoretical constraints that may exist on $m_t$, irrespective of any
phenomenological biases.} but it definitely leads to new physical consequences
which are not part of the usual weakly interacting Standard Model.

We are thus led to find an upper bound on $m_t\leftrightarrow\lambda_t$
such that if it is violated, we would say that the Standard Model has become
a strongly interacting theory in the Yukawa sector.\footnote{An analogous
strongly interacting Standard Model in the Higgs sector is obtained for
heavy Higgs boson masses. Corresponding upper bounds on $M_H$ can be obtained,
as described in Ref.~\cite{higgs,DJL}.} We will use the concept of partial-wave
unitarity to address this question. This is an arcane subject that is of
general applicability, but which is probably unfamiliar to present-day
students.

Consider two-body to two-body (\ie, $2\to2$) scattering processes. These
are called ``elastic" processes, as opposed to the $2\to n$ ($n>2$) inelastic
processes. The elastic $2\to2$ amplitudes can be decomposed in a partial-wave
expansion, which for scalar particles in the high-energy limit is given by
\begin{equation}
a_j^{2\to2}(s)=\coeff{1}{32\pi}\int_{-1}^{1} d\cos\theta\,
T(s,\cos\theta)P_j(\cos\theta),
\label{partial_waves}
\end{equation}
where $T(s,\cos\theta)$ are the usual Feynman amplitudes, $\theta$ is the
scattering angle, and $P_j$ are the Legendre polynomials. These partial-wave
amplitudes are related to the partial-wave projections of the $S$ matrix by
\begin{equation}
a_j^{2\to2}=\coeff{1}{2i}(S^{2\to2}_j-1).
\label{a_S}
\end{equation}
For inelastic processes we have instead $a_j^{2\to n}=S^{2\to n}_j/2i,\ n>2$
(\ie, no ``$-1$" since one necessarily has particle production). The unitarity
of the $S$ matrix (or of any of its partial waves $S_j$), $SS^\dagger=1$
implies that (summing over a row of $S$)
\begin{eqnarray}
1&=&\sum_b S_{ab}S^\dagger_{ba}=\sum_b |S_{ab}|^2\nonumber\\
&=&|S^{2\to 2}_j|^2+\sum_{n>2}|S^{2\to n}_j|^2\nonumber\\
&=&|2ia^{2\to 2}_j+1|^2+\sum_{n>2}|2ia^{2\to n}_j|^2\ .\nonumber
\end{eqnarray}
Or, as is it usually written
\begin{equation}
|a^{2\to2}_j-\coeff{i}{2}|^2+\sum_{n>2}|a^{2\to n}_j|^2=\coeff{1}{4}\ .
\label{unitarity}
\end{equation}
That is, the elastic amplitude $a^{2\to2}_j$ must lie on a circle (the
``unitarity circle") of radius $\eta/2$ centered at $(0,{1\over2})$ in the
complex plane, with  $\eta^2=1-4\sum_{n>2}|a^{2\to n}_j|^2$. This gives the
{\em Argand diagram} for the scattering amplitude, as shown below for the case
where inelastic processes are neglected (\ie, $\eta=1$).
\vspace{0.5cm}

\hrule
\vspace{0.2cm}
\noindent{\large\bf The Unitarity Circle}
\begin{figure}[h]
\vspace{2.2in}
\includegraphics{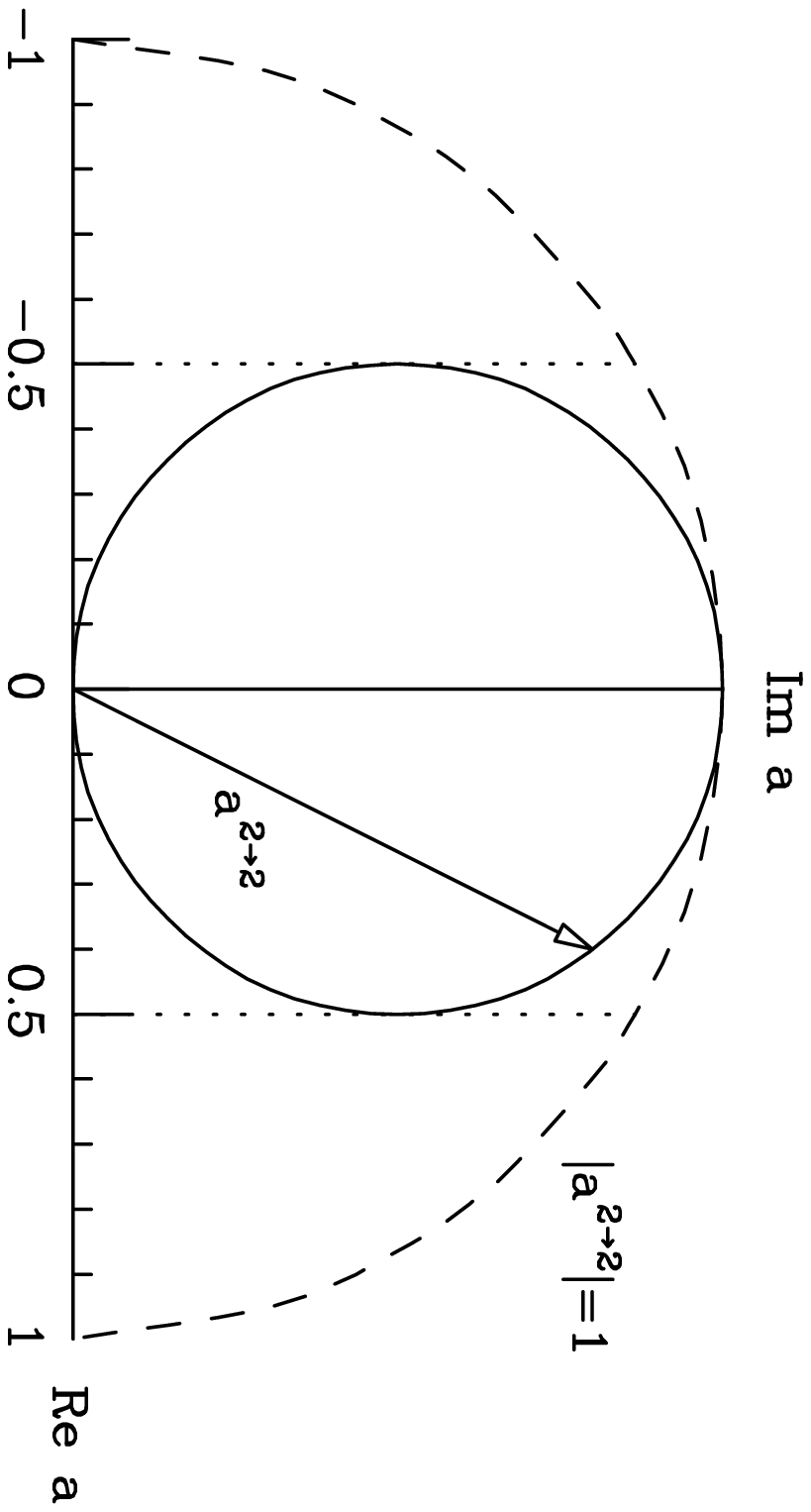}
\end{figure}
\vspace{0.5cm}
\hrule
\vspace{0.5cm}

The partial-wave amplitudes $a^{2\to2}_j$ can be calculated in perturbation
theory for any desired process. (Below we will be interested in $t\bar t\to
t\bar t$ scattering.) However, it is the all-orders amplitude that must lie on
the unitarity circle. The idea of ``unitarity bounds" is that if a given
partial-wave amplitude (usually $j=0$), calculated to some low-order in
perturbation theory (usually tree-level), does not lie on the unitarity circle,
then it must be that higher-loop corrections are important in order to restore
unitarity -- \ie, there is a breakdown of perturbation theory.

Without thinking too much about the unitarity circle, it is common practice
to demand $|a^{2\to2}_j|\le1$ (\ie, to be inside the dashed semi-circle)
or ${\rm Re}\,|a^{2\to2}_j|\le{1\over2}$ (to be between the dotted vertical
lines), which are clearly necessary conditions but may not be sufficient.
Taking advantage of the unitarity circle requires a one-loop calculation
of the partial-wave amplitudes, since at tree-level all amplitudes are
real. The above discussion indicates that if one is limited to a tree-level
calculation, then a more realistic unitarity constraint would be  ${\rm
Re}\,|a^{2\to2}_j|\ll{1\over2}$, and perhaps  ${\rm
Re}\,|a^{2\to2}_j|\le{1\over4}$ is a sensible educated guess of the actual
one-loop constraint.\footnote{One-loop unitarity analyses in the Higgs sector
support this expectation \cite{DJL}.}

In the Standard Model we are interested in constraining the top-quark Yukawa
coupling. Therefore we need to consider processes which depend strongly
on $\lambda_t$, such as $t\bar t\to t\bar t$ (which proceeds through $s$- and
$t$-channel Higgs-boson and $Z$ exchanges). In the high-energy
limit\footnote{When considering the high-energy limit, only coefficients of
dimension-four operators in the Lagrangian (such as Yukawa couplings) will
survive. Mass parameters (\eg, soft-supersymmetry-breaking parameters)
cannot be bounded in this fashion.}
(\ie, $s\gg m^2_t,m^2_H$) this amplitude is given by $T=-3\lambda^2_t$
\cite{CFH}. The $j=0$ partial wave follows from Eq.~(\ref{partial_waves}):
$a_0=-3\lambda^2_t/16\pi$. The ``unitarity bound" arising from $|{\rm
Re}\,a_0|\le{1\over2}$ is then
\begin{equation}
\lambda_t\le\sqrt{8\pi\over3}\approx2.89\ .
\label{SMlimit}
\end{equation}
Requiring instead $|{\rm Re}\,a_0|\le{1\over4}$ as our ``one-loop" guess, gives
$\lambda_t\le\sqrt{4\pi\over3}\approx2.05$. From Eq.~(\ref{mtSM}) we then
obtain an upper bound on $m_t$
\begin{equation}
m_t=\lambda_t {v_0\over\sqrt{2}}<\lambda^{\rm max}_t
{v_0\over\sqrt{2}}\approx500\GeV\qquad {\rm (Standard\ Model)},
\label{SMmtbound}
\end{equation}
or $m_t\lsim355\GeV$ using the ``one-loop" bound.

Within the Standard Model there is another theoretical constraint on $m_t$
that one can find by requiring stability of the minimum of the Higgs potential
(``vacuum stability"). The one-loop effective potential is given by \cite{Sher}
\begin{equation}
V=V_0+{B\over64\pi^2}\,\phi^4\ln{\phi^2\over Q^2},
\end{equation}
with $B=(6m^4_W+3m^4_Z-12m^4_t)/v^4$. If $m_t$ is too large, $B$ will turn
negative and the potential becomes unbounded from below. For large values of
$\phi$, the expansion parameter is enhanced by large logarithms, and one must
use renormalization group methods to obtain a reliable result. For a given
upper bound on the mass scale where the Standard Model breaks down, there
is a constraint in the $m_t-m_H$ plane \cite{LSZ}
\begin{equation}
{\rm For}\ \phi\lsim1\TeV\quad \Rightarrow \quad m_t\lsim2 m_H+60\GeV\ .
\label{stab}
\end{equation}
This constraint is not very useful, although it can become strictier
is the scale of new physics is pushed up.

\section{Grand Unified Theories (GUTs)}

For our purposes it will suffice to assume that the Standard Model is valid
up to scales $M_U={\cal O}(10^{15}\GeV)$ where the new GUT physics comes in.
The constraint derived in Eq.~(\ref{SMlimit}) for a weakly interacting theory
remains valid, except that now it must be satisfied at all mass scales $Q$,
\ie,
\begin{equation}
\lambda_t(Q)\lsim2.89,\qquad M_Z\lsim Q\lsim M_U\ .
\label{GUTslimit}
\end{equation}
Our weakly interacting assumption required some justification within the
Standard Model, whereas in GUTs this assumption is usually (and tacitly)
made in order to be able to connect the low- and high-mass scales via the
(one-loop) renormalization group equations (RGEs). In fact, in order to
enforce the constraint in Eq.~(\ref{GUTslimit}) we need to use the RGEs
for $\lambda_t$ and the gauge couplings (valid for $M_Z\lsim Q\lsim M_U$)
\begin{eqnarray}
{d\lambda_t\over dt}&=&{\lambda_t\over8\pi^2}\left[a\lambda^2_t-\sum_i
c_ig^2_i\right]\\
{dg_i\over dt}&=&{b_i\over16\pi^2}\,g^3_i
\end{eqnarray}
where the various coefficients depend on the theory at hand, \ie, the Standard
Model or its supersymmetric version. These coefficients are given below.
\vspace{0.5cm}

\hrule
\begin{center}
\begin{tabular}{|c|c|c|c|c|c|c|c|}\hline
&$a$&$c_1$&$c_2$&$c_3$&$b_1$&$b_2$&$b_3$\\ \hline
SM&$9/4$&$17/40$&$9/8$&$4$&$41/10$&$-19/6$&$-7$\\ \hline
SUSY&$3$&$13/30$&$3/2$&$8/3$&$33/5$&$1$&$-3$\\ \hline
\end{tabular}
\end{center}
\hrule
\vspace{0.5cm}

If all other Yukawa couplings can be neglected (as is the case in the Standard
Model), then the above RGEs can be solved analytically. We obtain \cite{DL}
\begin{equation}
\lambda_t^2(t)={\lambda^2(0)F(t)\over 1-{a\over4\pi^2}\lambda^2(0)T(t)}\ ,
\label{yuk}
\end{equation}
where $t=\ln(Q/M_Z)$ and $T(t)=\int_0^t F(t')dt'$
with
\begin{equation}
F(t)=\prod_{i=1}^3\left[1-{b_i\over8\pi^2}\,g^2_i(0)t\right]^{2c_i/b_i}\ .
\end{equation}
To evaluate these expressions we use:
\begin{eqnarray}
g^2_1(0)&=&\left({5\over3}\right){4\pi\alpha_e\over\cos^2\theta_W}=0.212,\\
g^2_2(0)&=&{4\pi\alpha_e\over\sin^2\theta_W}=0.426,\\
g^2_3(0)&=&4\pi\alpha_3=1.508,
\end{eqnarray}
with\footnote{To be consistent with one-loop gauge coupling unification, we
should use the predicted value of $\sin^2\theta_W$, \ie,
$\sin^2\theta_W=0.2+(7/15)(\alpha_e/\alpha_3)$.}
\begin{equation}
\alpha_e^{-1}=127.9,\quad \sin^2\theta_W=0.2304,\quad \alpha_3=0.120\ .
\end{equation}

The calculated values of $\lambda_t(Q)$ versus the scale $Q$ are shown in
the following plot. These numbers only depend on the assumed value of
$\lambda_t(0)=\lambda_t(Q=M_Z)$. The horizontal dashed line represents
the constraint in Eq.~(\ref{GUTslimit}).
\clearpage

\begin{figure}[t]
\vspace{3.5in}
\includegraphics{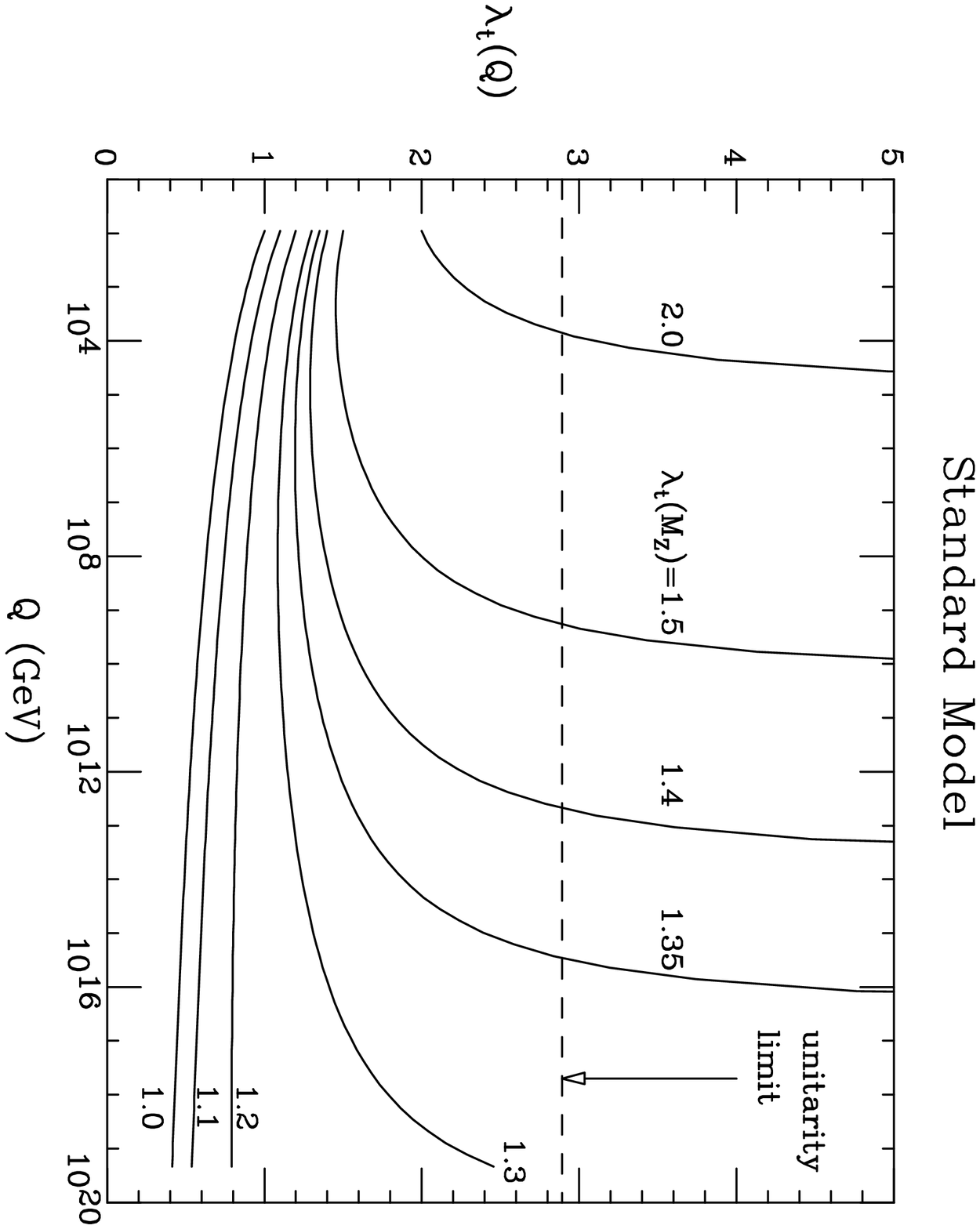}
\vspace{2cm}
\end{figure}
\hrule
\vspace{0.5cm}

Note that as $\lambda_t(0)$ increases, the unitarity limit (dashed line) is
crossed for lower and lower values of $Q$: for $\lambda_t(0)=1.4\,(1.5)$,
$Q^{\rm max}\sim10^{12}\,(10^9)\GeV$. This is the maximum scale for which such
theory remains weakly interacting. If we want $Q^{\rm max}>M_U=10^{15}\GeV$,
then there is an upper bound on $\lambda_t(0)$:
\begin{equation}
\lambda_t^{\rm max}(0)\approx1.357\ .
\label{GUTs0limit}
\end{equation}
{}From Eq.~(\ref{mtSM}) we then get the corresponding upper bound on $m_t$
\begin{equation}
m_t<\lambda_t^{\rm max}(0){v_0\over\sqrt{2}}\approx236\GeV\qquad {\rm
(GUTs-unitarity)}\,.
\label{GUTsmtbound}
\end{equation}
This result is to be compared with that for the Standard Model without any
``RGE improvement" (in Eq.~(\ref{SMmtbound})): the value of $\lambda^{\rm
max}_t$ is reduced from 2.89 down to 1.36, \ie, by more than a factor of two!
Using the ``one-loop" value of 2.05 instead of 2.89, reduces
the 1.357 result to 1.342. Thus, a 30\% reduction at the low scale only
amounts to a 1\% reduction at the high scale.

Note that given a sufficiently high mass scale, the Yukawa coupling will {\em
always} blow up (a ``Landau pole"), \ie, the denominator of Eq.~(\ref{yuk})
becomes zero for $T(t_c)=4\pi^2/a\lambda^2(0)$ which determines the critical
scale $t_c$ for {\em any} value of $\lambda(0)$.\footnote{In practice,
$Q_c=M_Ze^{t_c}$ exceeds the Planck scale for $\lambda(0)<1.3$.} This behavior
of the theory is called
{\em triviality} \cite{triviality}, since the only allowed value of the
$\lambda(0)$ parameter appears to be zero, \ie, the theory is trivial. This
result would be worrisome if the theory had no {\em cutoff}, that is, no
scale beyond which the theory changes and the previous reasoning does not
apply. However, all possible theories of this kind have a natural cutoff
at the Planck scale: no matter what happens at the GUT scale or
beyond, the theory must change at the Planck scale to accomodate the
gravitational interactions. Moreover, the change which is needed is not
just into another gauge theory, since the same problem would reappear
above the cutoff scale. In the case of string theory, the theory becomes
finite and the parameters do not run anymore. Therefore, a theory of
extended objects (like string theory) appears to be also motivated as a final
solution to the triviality problem of the Yukawa sector.

In this connection, instead of demanding that the GUT theory be weakly
interacting all the way up the unification scale, one can simply restrict the
theory such that the Yukawa coupling does not blow up before the unification
scale \cite{MPP}. So the question now is: for what $\lambda_t^{\rm max}(0)$
is $\lambda_t(Q)<\infty$ for $Q=M_U$? We find
\begin{equation}
\lambda_t^{\rm max}(0)\approx1.373\ ,
\label{lambdat-triviality}
\end{equation}
and therefore
\begin{equation}
m_t<\lambda_t^{\rm
max}(0){v_0\over\sqrt{2}}\approx239\GeV\qquad{\rm(GUTs-triviality)}\,.
\label{trivialitytbound}
\end{equation}

Note that there is little difference between the $m_t$ upper bounds with the
unitarity (\ref{GUTsmtbound}) or the triviality (\ref{trivialitytbound})
requirements. The reason is that large values of Yukawa coupling at the high
scale are driven at low energies to an ``infrared fixed point" which depends on
the scale where $\lambda_t(Q)\to\infty$, \ie, for $Q=10^{15}\GeV$, the fixed
point is $\lambda_t(0)=1.373$.

Finally, we can re-consider the constraint coming from the stability of the
Higgs vacuum, but this time allowing the Higgs field to take values as large
as the unification scale. The result is \cite{LSZ}
\begin{equation}
m_t\lsim\coeff{1}{2}m_H+100\GeV,
\end{equation}
which is much stronger than the one obtained in Eq.~(\ref{stab}) because of the
``RGE strengthening" of the limit when allowing higher mass scales.

\section{Supersymmetric unified theories (SUSY GUTs)}
The introduction of supersymmetry into GUTs is basically required to make
sense of the gauge hierarchy problem. The analysis performed above for the
case of (non-supersymmetric) GUTs has a direct parallel in the case of
SUSY GUTs, except for a crucial difference: in a supersymmetric theory one
must have at least two Higgs doublets in order to have Yukawa couplings
(and therefore masses) for both up-type and down-type quarks. In the minimal
case (which we will focus on) the Higgs sector consists of only two Higgs
doublets ($H_1,H_2$), and therefore there are two vacuum expectation values
$v_1,v_2$. The sums of the squares of these is constrained (\ie,
$v_1^2+v_2^2=v^2_0$) but their ratio ($\tan\beta=v_2/v_1$) is not. The fermion
Yukawa couplings then depend on this new parameter:
\begin{eqnarray}
m_t&=&\lambda_t {v_0\over\sqrt{2}}\sin\beta,\\
m_b&=&\lambda_b {v_0\over\sqrt{2}}\cos\beta,\\
m_\tau&=&\lambda_\tau {v_0\over\sqrt{2}}\cos\beta.
\end{eqnarray}
When $\tan\beta$ is large ($\gsim40$) the Yukawa couplings of the bottom
quark and tau lepton are enhanced, and one should include these also in
the RGE scaling.

The partial-wave unitarity analysis performed for the Standard Model is also
applicable to a low-energy supersymmetric theory, as follows. In the
supersymmetric theory one has the same
\begin{equation}
t\bar t\to \gamma,Z\to t\bar t
\label{tt}
\end{equation}
amplitude (since supersymmetric particles cannot be exchanged in tree-level
diagrams when $R$-parity is conserved). In addition, one has amplitudes
involving the superpartners to the top quarks (the top-squarks $\tilde t$)
\begin{equation}
\tilde t\tilde t\to\gamma,Z\to \tilde t\tilde t,
\label{stst}
\end{equation}
and amplitudes involving superpartners of the $\gamma,Z$ (the neutralinos
$\chi^0_i$)
\begin{equation}
t\tilde t\to\chi^0_i\to t\tilde t.
\end{equation}
There are also ``crossed" amplitudes, like $t\gamma\to t\to t\gamma$ and
$t\chi^0_i\to \tilde t\to t\chi^0_i$, etc. All of these amplitudes are
proportional to $\lambda^2_t$ and should be taken into account in obtaining
the unitarity limit. A great simplification occurs when one considers the
high-energy limit, in which all mass parameters decouple and only the
$\lambda_t$ dependence survives \cite{DL}. In this case supersymmetry is
effectively restored (since all soft-supersymmetry-breaking parameters become
negligible) and the various amplitudes get correlated.
The unitarity constraint $|{\rm Re}\,a_0|\le{1\over2}$ applies to all
eigenvalues of the matrix of coupled channels. This coupling usually results
in a strengthening of the unitarity limit by a factor of order 1. If one
nonetheless neglects the channel coupling and only considers the channels in
Eqs.~(\ref{tt},\ref{stst}), the result is the same as in the Standard Model
case (\ie, $\lambda_t<\sqrt{8\pi/3}$) since the effectively restored
supersymmetry entails same amplitudes for these two channels.

Despite the bound on $\lambda_t$ being the same as in the Standard Model case,
the corresponding upper bound on $m_t$ can be stronger because of the
$\tan\beta$ dependence
\begin{equation}
 m_t<\lambda_t^{\rm max}\cdot174\cdot\sin\beta\approx500\,\sin\beta\,\GeV
\qquad{\rm(SUSY)}\,.
\end{equation}
For $\tan\beta=1.0,1.3,1.5,2,3$ we get $m_t\lsim355,400,420,450,475\GeV$.

On the other hand, in a weakly interacting SUSY GUT, the bound
$\lambda_t\lsim2.89$ must hold for all scales up to the unification scale
$M_U\sim10^{16}\GeV$ in this case. We can repeat the exercise for the GUT
case and run the supersymmetric RGEs for the gauge and Yukawa couplings.
Neglecting the $\lambda_b$ and $\lambda_\tau$ contributions (for not too
large values of $\tan\beta$) we find for the running top-quark Yukawa
coupling the following.
\vspace{0.5cm}
\hrule
\begin{figure}[h]
\vspace{3.5in}
\includegraphics{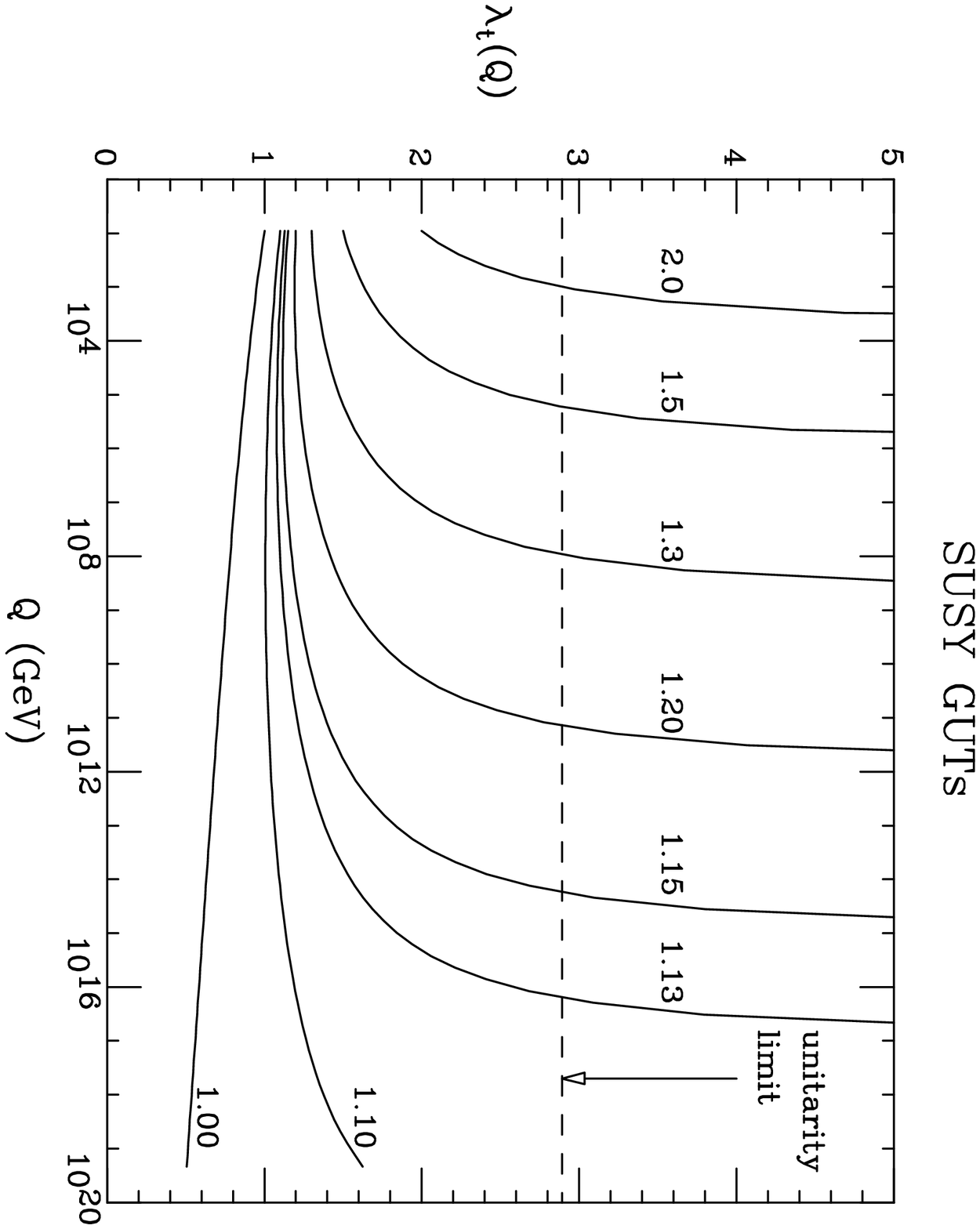}
\vspace{2.5cm}
\end{figure}
\hrule
\vspace{0.5cm}

Note that for increasingly larger values of $\lambda_t(0)$, the critical
scale $Q_c$ where the unitarity limit is crossed, decreases. Requiring that
this scale be above the unification scale \cite{DL} ($M_U\sim10^{16}\GeV$)
gives the upper limit on $\lambda_t(0)$
\begin{equation}
\lambda_t^{\rm max}(0)\approx1.132\ ,
\end{equation}
which entails the following upper bound on $m_t$
\begin{equation}
m_t<\lambda_t^{\rm max}(0){v_0\over\sqrt{2}}\sin\beta\approx197\,\sin\beta\GeV
\qquad {\rm (SUSY\ GUTs-unitarity)}\,.
\end{equation}
In this case the ``RGE improvement" decreases the maximum value of $\lambda_t$
by 60\%, making the upper bound on $m_t$ that much stronger.

As in the GUTs case, we can also determine the critical value of $\lambda_t(0)$
above which $\lambda_t(Q)$ would blow up before the unification scale. We
find
\begin{equation}
\lambda_t^{\rm max}(0)\approx1.138\ ,
\end{equation}
which entails the following upper bound on $m_t$
\begin{equation}
m_t<\lambda_t^{\rm max}(0){v_0\over\sqrt{2}}\sin\beta\approx198\,\sin\beta\GeV
\qquad {\rm (SUSY\ GUTs-triviality)}\,.
\end{equation}

The above results are sensitive to various ``thresholds" encountered between
$M_Z$ and $M_U$, as well as the value of $\alpha_3$, and the contribution
from $\lambda_b$. A more refined calculation gives the same qualitative
results, but slightly stronger bounds \cite{aspects}
\begin{equation}
m_t\lsim190\sin\beta\GeV.
\label{refined}
\end{equation}
This upper bound on $m_t$ was of ``academic" interest when originally
derived \cite{DLplb}, but has since become very relevant given the experimental
trend towards increasingly heavier top-quark masses since then.\\

\noindent{\bf Subtlety}. The above-quoted top-quark masses are the ``running"
masses at the scale $Q=m_t$, \ie, $m_t(m_t)$. On the other hand, the
experimentally observable top-quark mass is the ``pole" mass, which is related
to the running mass through \cite{GBGS}
\begin{equation}
m_t^{\rm pole}=m_t\left[1+\coeff{4}{3}{\alpha_s(m_t)\over\pi}
+K_t\left({\alpha_s(m_t)\over\pi}\right)^2\right]
\end{equation}
where
\begin{equation}
K_t=16.11-1.04\sum_{m_{q_i}<m_t}\left(1-{m_{q_i}\over m_t}\right)\approx11.
\end{equation}
We thus obtain
\begin{equation}
m^{\rm pole}_t\approx 1.067 m_t\ .
\end{equation}
Distinguishing between these top-quark masses is unnecessary
at lowest order, but becomes essential in a next-to-leading order calculation.
Equation~(\ref{refined}) then gives
\begin{eqnarray}
m_t^{\rm pole}&\lsim&143\GeV\qquad{\rm for}\ \tan\beta=1\label{ua}\\
&\lsim&169\GeV\qquad{\rm for}\ \tan\beta=1.5\\
&\lsim&181\GeV\qquad{\rm for}\ \tan\beta=2\\
&\lsim&198\GeV\qquad{\rm for}\ \tan\beta=5\\
&\lsim&202\GeV\qquad{\rm for}\ \tan\beta=10\label{ue}
\end{eqnarray}
Note that $\tan\beta=1$ is almost ruled out experimentally: at the 95\% C.L.
the CDF result requires $m_t\gsim141\GeV$ and the global fit requires
$m_t\gsim144\GeV$. Should the top-quark mass actually be $m_t=162\,(174)\GeV$,
we would deduce that $\tan\beta>1.33\,(1.67)$.

The fact that the unitarity and triviality SUSY GUTs (or GUTs) upper bounds on
$\lambda_t$ and $m_t$ are so close is related to an {\em infrared fixed point}
in the RGE for the Yukawa coupling. We can see this phenomenon clearly if
we plot the low-energy value of the Yukawa coupling $\lambda_t(M_Z)$ which is
obtained for a given high-energy value $\lambda_t(M_U)$:
\vspace{0.5cm}
\hrule
\begin{figure}[h]
\vspace{3.5in}
\includegraphics{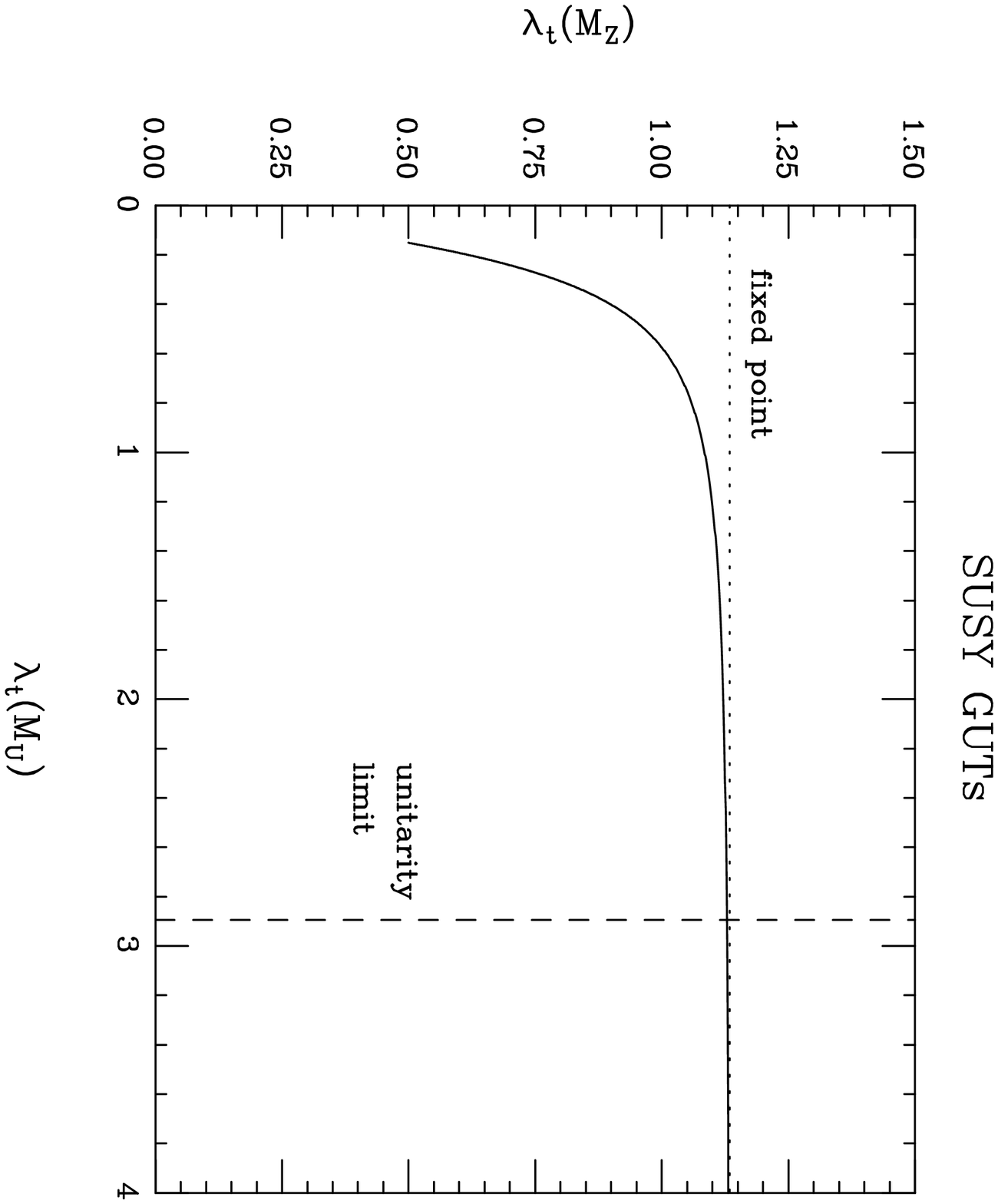}
\vspace{3.3cm}
\end{figure}
\hrule
\vspace{0.5cm}

The picture is clear: for sufficiently large values of the high-energy
Yukawa coupling, its low-energy counterpart will be driven to the same
``fixed point value". This phenomenon explains why it does not matter what we
choose for $\lambda_t(M_U)$ if it exceeds $\sim1.5$. Thus the little
difference between the unitarity and triviality constraints. Also, since
the tree-level unitarity constraint is not very precise, as discussed above,
one could consider a ``one-loop" guess for it by taking half of its value.
This brings 2.89 down to 2.05, which gives a very similar value of
$\lambda_t(M_Z)$, as the figure shows. If one decides that the fixed point
value of $\lambda_t$ is the preferred one, then one obtains the ``fixed point
value" of the top-quark mass, \ie, $m^{\rm pole}_t\approx(203\GeV)\sin\beta$,
and the upper bounds in Eqs.~(\ref{ua})--(\ref{ue}) are actually attained.

Further constraints on $m_t$ can be obtained by assuming special relations
among the Yukawa couplings at the unification scale.

\begin{itemize}
\item $SU(5)$-like: $\lambda_b=\lambda_\tau$ at $M_U$ gives a definite curve in
the $(m_t,\tan\beta)$ plane \cite{su5}.
\item $SO(10)$-like: $\lambda_t=\lambda_b=\lambda_\tau$ at $M_U$ gives a
point in the $(m_t,\tan\beta)$ plane \cite{so10}.
\end{itemize}
Both relations are quite sensitive to the choice of $m_b$ and $\alpha_3$,
and also depend on the threshold structure at the GUT and electroweak scales.
Below we show typical predictions from these relations (data from
Refs.~\cite{Kane,Carena}). Note that the $SO(10)$-like relations require large
values of $\tan\beta$, whereas the $SU(5)$-like relation prefers small
($\sim1$) or large values; intermediate values are allowed only for large
values of $m_t$.
\vspace{0.5cm}
\hrule
\begin{figure}[h]
\vspace{3.5in}
\includegraphics{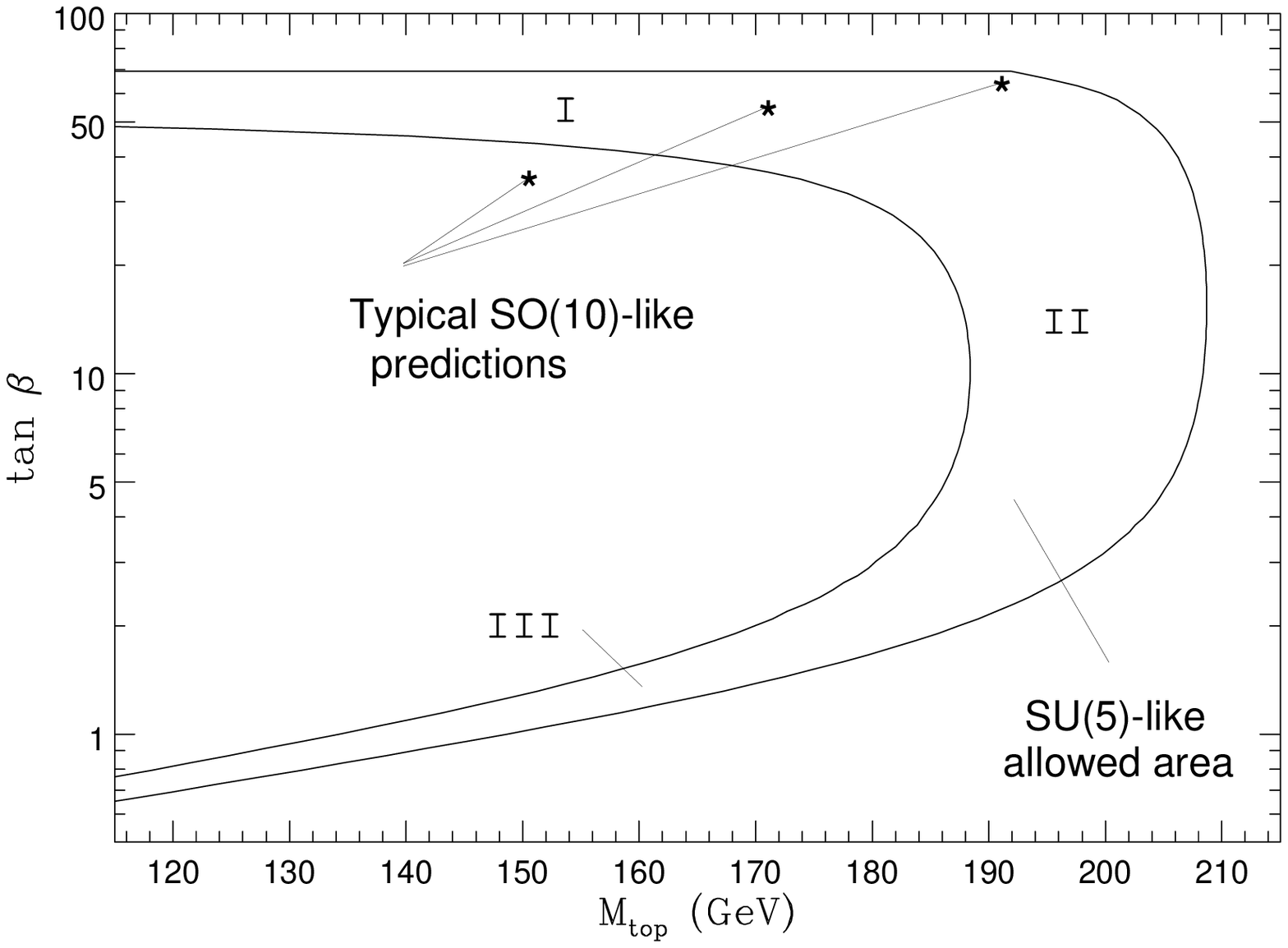}
\end{figure}
\vspace{2.5cm}
\hrule

\section{Supergravity}
Local supersymmetry or supergravity allows the spontaneous breaking of
supersymmetry via the super-Higgs effect, and thus the calculation of the
soft-supersymmetry-breaking parameters in terms of a few input functions
(the superpotential, the K\"ahler potential, and the gauge kinetic function).
Usually one assumes that the resulting soft-supersymmetry-breaking parameters
are universal at the unification scale. The running of the various scalar
and gaugino masses  from the unification scale down to the electroweak scale is
governed by a set of coupled RGEs. Of particular relevance are the squared
Higgs-doublet masses, one of which can turn negative at sufficiently low
scales, signaling the breaking of the electroweak symmetry by radiative effects
-- {\em the radiative electroweak breaking mechanism} \cite{EWx,LN}.

Consider one of these RGEs schematically
\begin{equation}
{d\widetilde m^2\over dt}={1\over(4\pi)^2}\left\{
-\sum_i c_i g^2_i M^2_i + a\lambda^2_t\left(\sum_i\widetilde
m^2_i\right)\right\},
\end{equation}
where $g_i$ are the gauge couplings, $M_i$ are the gaugino masses, and
$c_i$ and $a$ are various positive coefficients. If $\lambda_t$ is ``not small"
compared to the gauge couplings, then $\widetilde m^2$ could turn negative at
low energies. For concreteness let's say $\lambda_t\gsim g\sim0.6$, which
implies
\begin{equation}
m_t=174\lambda_t\sin\beta\gsim100\sin\beta\gsim75\GeV.
\end{equation}
This qualitative result is what people have in mind when stating that
radiative breaking works only for ``heavy top quarks". Heavy here means heavier
than the prevailing prejudice a decade ago, \ie, $m_t\gg m_b$. All imaginable
top-quark masses today should satisfy this constraint automatically.

Nonetheless, radiative electroweak breaking does impose constraints in the
$(m_t,\tan\beta)$ plane \cite{aspects}, as the following figure shows.
The allowed region is bounded completely:
\begin{itemize}
\item Top boundary: the vacuum of the Higgs potential is untable above this
line. This in effect is the analogue of the vacuum stability constraint
discussed in connection with the Standard Model.
\item Upper corner: this corresponds to a fixed point in $\lambda_b$, which
translates into an upper bound on $\tan\beta$ since $m_b\propto
\lambda_b\cos\beta$.
\item Right boundary: the fixed point in $\lambda_t$ produces an upper bound on
$m_t$ as a function of $\tan\beta$, as discussed above.
\item Bottom boundary: $\tan\beta>1$ is required by radiative electroweak
symmetry breaking.
\item Left boundary: the radiative breaking mechanism does not work to the left
of this line, \ie, $\mu^2<0$ would be required.
\end{itemize}
As the figure shows, there is always a minimum value of $m_t$ which is
required. Note that for $\mu>0$, it may be possible to reach small values of
$m_t$ for values of $\tan\beta$ sufficiently close to 1.\footnote{I would like
to thank Chris Kolda for pointing this out to me.} However, such region of
parameter space is highly disfavored since the tree-level contribution to the
lightest Higgs-boson mass nearly vanishes and the one-loop contribution is
small because of the small values of $m_t$. In the figure one can also
appreciate the effect of using the one-loop effective potential in determining
the value of $\mu$: the largest deviation from the tree-level result occurs
at the left boundary.

\begin{figure}[t]
\vspace{5in}
\includegraphics{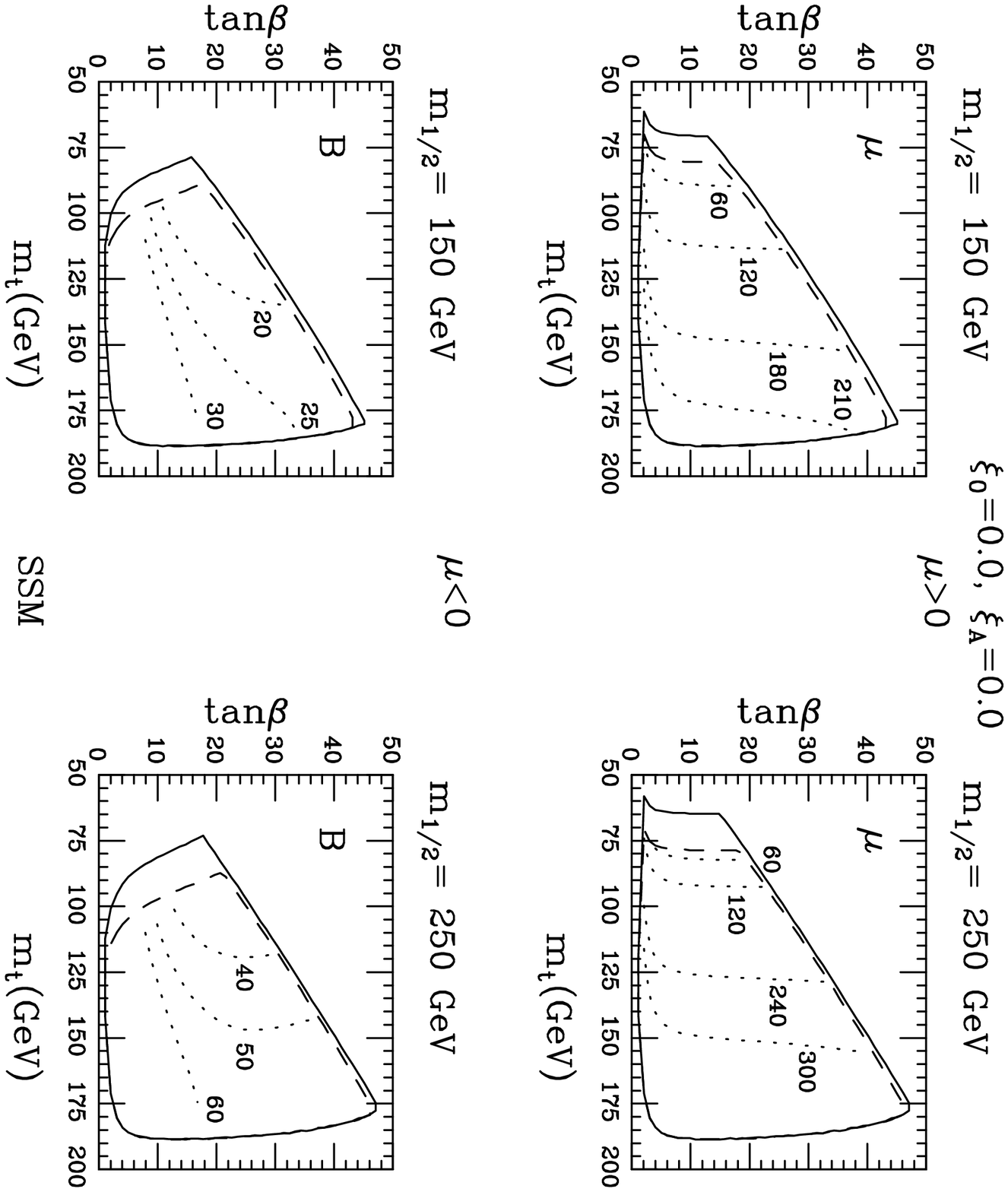}
\vspace{1cm}
\end{figure}
\vspace{-3cm}
\hrule
\vspace{3cm}

\section{Superstrings}
Superstrings provide the only known consistent theory of quantum gravity.
They also provide means for calculating all parameters in a string model in
terms of dimensionless constants or dynamically determined parameters.
In particular, the Yukawa couplings can be calculated in a given string
``vacuum" in terms of the string gauge coupling and possibly the expectaction
values of some ``moduli" fields which parametrize deformations of the chosen
vacuum. This calculational property should be regarded as unique and
fundamental as that of finite quantum gravitational interactions.

In the free-fermionic formulation of the heterotic string in four
dimensions \cite{FFF}, a typical result of such calculations of Yukawa
couplings is \cite{revamped,decisive}
\begin{equation}
\lambda_t(M_U)=\sqrt{2}g\cos\theta_t\,,
\end{equation}
where $\cos\theta_t$ is an ``effective" coefficient which may arise from some
mixing of states leading to the physical states, or is simply another
coefficient which appears in the actual calculation. Typical values of this
effective coefficient are $\cos\theta_t=1,{1\over\sqrt{2}},{1\over2}$.
Therefore, the typical string predictions for $\lambda_t$ are not small, but
since $\lambda_t=\sqrt{2}g\cos\theta_t\lsim1$, these are always below the
unitarity limit of 2.89. Moreover, this prediction is also generically not
small, therefore large values of $m_t$ are typical predictions of string
models. In the following figure we show the calculated values of
$\lambda_t(M_U)$ for given (running) top-quark masses and fixed values of
$\tan\beta$ \cite{tpaper}. Two typical string predictions are denoted by
horizontal dashed lines. Note that one expects $m_t\sim160-185\GeV$
(or $m^{\rm pole}_t\sim170-195\GeV$).

\vspace{0.25cm}
\hrule
\vspace{0.2cm}
\begin{figure}[h]
\vspace{4.5in}
\includegraphics{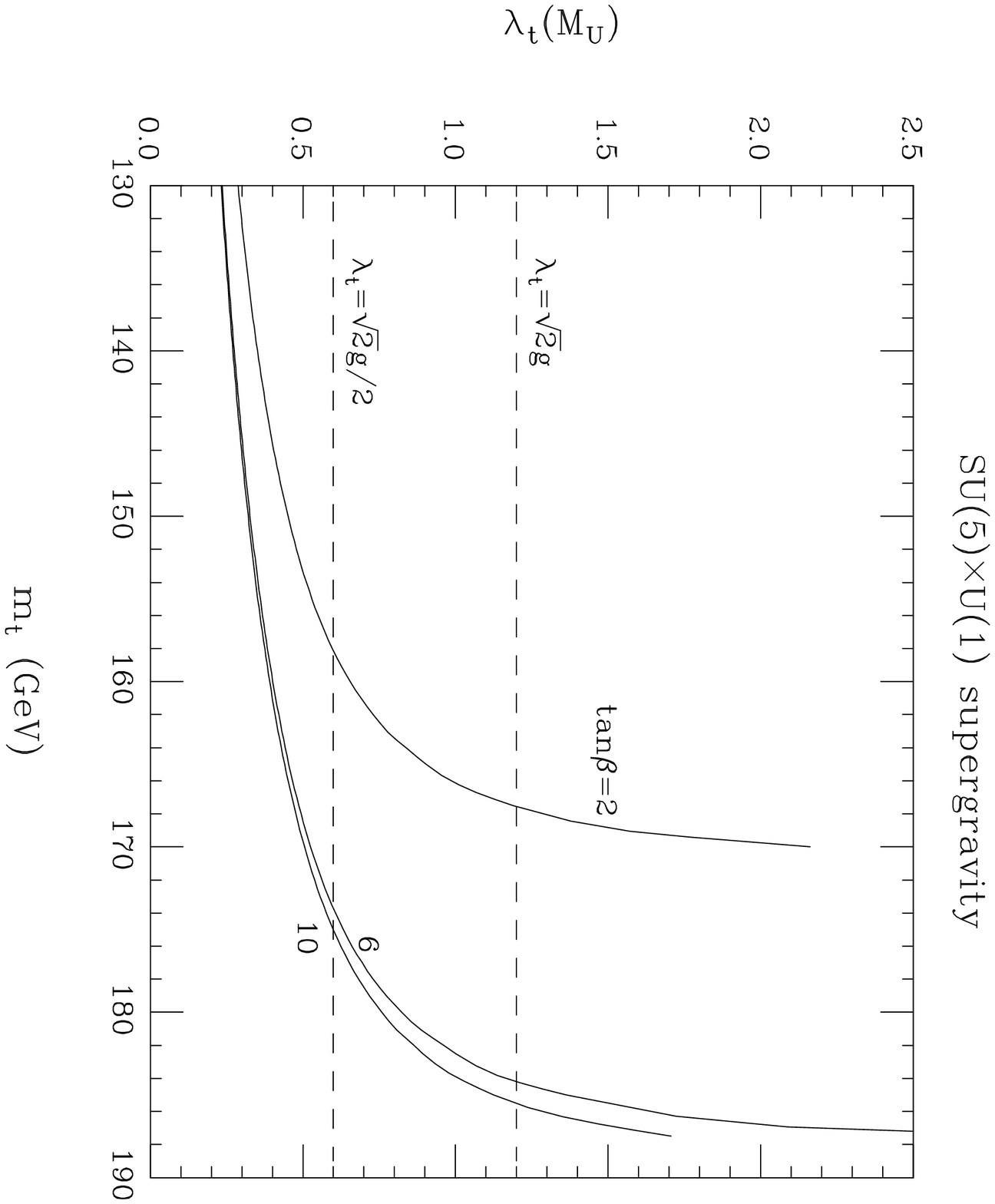}
\end{figure}
\clearpage

\section{Conclusions}
In an attempt to make some sense of the mounting evidence for the top-quark
mass, we have recalled the various theoretical constraints that should be
satisfied in a sensible theory. These constraints depend on the assumptions
made about the theory, \ie, the energy range where it is expected to hold,
its matter content, the larger theory which may embed it, etc.

Most of the constraints which we derived were based on the assumption that the
theory remained weakly interacting in its presumed regime of validity. This
is an assumption that does not need to hold, \ie, the theory could have a
strongly interacting phase with new physical predictions. However, if one
decides to ignore this possibility (as it is tacitly done all the time), then
one must be consistent in requiring that no sector of the theory violates
this tacit assumption.

The requirement of weakly interacting theories results in upper bounds on
$m_t$:
\begin{itemize}
\item Standard Model: $m_t\lsim500\GeV$;
\item SUSY: $m_t\lsim500\,\sin\beta\GeV$;
\item GUTs: $m_t\lsim240\GeV$;
\item SUSY GUTs: $m_t\lsim200\sin\beta\GeV$.
\end{itemize}
On the other hand, radiative electroweak symmetry breaking requires a
not-too-small value of $m_t$, typically $m_t\gsim75\GeV$.
We also discussed {\em predictions} for the top-quark mass in superstring
models based on the gauge group $SU(5)\times U(1)$. These predictions satisfy
all the theoretical constraints discussed previously and typically require
$m^{\rm pole}_t\sim170-195\GeV$.

All of these expectations are based solely on theoretical concepts. On the
other hand, phenomenological expectations appear to favor
$m_t\sim160\pm10\GeV$. We therefore conclude that SUSY GUTs expectations and
superstring predictions are in good agreement with present experimental
expectations. Moreover, in these theoretical frameworks one can {\em naturally}
understand the apparently large value of the top-quark mass.


\begin{thebibliography}{99}
\bibitem{CDF} The CDF Collaboration, ``Evidence for top quark production in
$\bar pp$ collisions at $\sqrt{s} = 1.8\TeV$", Fermilab-Pub-94/097-E (April
1994) and \PRL{73}{94}{225}.
\bibitem{LEPmt}The LEP Collaborations ALEPH, DELPHI, L3 and OPAL,
\PLB{276}{92}{247}; The LEP Collaborations ALEPH, DELPHI, L3 and OPAL, and the
LEP Electroweak Working Group, CERN preprint CERN/PPE/93-157.
\bibitem{EFLmt}See \eg, J.~Ellis, G.L.~Fogli and E.~Lisi, \PLB{292}{92}{427};
G.~Altarelli, R.~Barbieri and F.~Caravaglios, \NPB{405}{94}{3};
A.~Blondel and C.~Verzegnassi, \PLB{311}{93}{346};
P.~Langacker, in {\it Recent Directions in Particle Theory: From
Superstrings to the Standard Model\/}, ed.~by J.~Harvey and J.~Polchinski
(World Scientific, Singapore, 1993), p.~141; G.~Montagna, O.~Nicrosini and
G.~Passarino, \PLB{303}{93}{170}; V.A.~Novikov, L.B.~Okun, A.N.~Rozanov and
M.I.~Vysotsky, CERN-TH.7217/94 (Apr.~1994); D.~Schaile, CERN-PPE/93-213
(Dec.~1993).
\bibitem{EFL} J.~Ellis, G.L.~Fogli and E.~Lisi,  CERN-TH.7261/94 (May 1994).
\bibitem{reviews} For recent reviews see \eg, J.L. Lopez, D.V. Nanopoulos, and
A. Zichichi, Nuovo Cimento Rivista, {\bf17}(1994) 1 and Prog. Part. Nucl. Phys.
{\bf 33} (1994) 303.
\bibitem{Bardeen}W. Bardeen, C. Hill, and M. Lindner, \PRD{41}{90}{1647}.
\bibitem{higgs}M. Veltmann, Acta Phys. Pol. B8 (1977) 475; C. Vayonakis,
Lett. Nuovo Cimento, {\bf17} (1976) 383;
B. W. Lee, C. Quigg, and H. B. Thacker, \PRD{16}{77}{1519};
M. L\"uscher and P. Weisz, \PLB{212}{89}{472}; W. Marciano, G. Valencia,
and S. Willenbrock, \PRD{40}{89}{1725}.
\bibitem{DJL} L. Durand, J. Johnson, and \JL, \PRL{64}{90}{1215} and
\PRD{45}{92}{3112}.
\bibitem{CFH} M. Chanowitz, M. Furman, and I. Hinchliffe, \NPB{153}{79}{402};
W. Marciano, G. Valencia, and S. Willenbrock as in Ref.~\cite{higgs}.
\bibitem{Sher}For a review see, M. Sher, \PRT{179}{89}{273}.
\bibitem{LSZ}M. Lindner, M. Sher, and H. W. Zagluer, \PLB{228}{89}{139}.
\bibitem{DL}L. Durand and \JL, \PRD{40}{89}{207}.
\bibitem{triviality}D. Callaway, \PRT{167}{88}{241}.
\bibitem{MPP}L. Maiani, G. Parisi, and R. Petronzio, \NPB{136}{78}{115};
N. Cabibbo, L. Maiani, G. Parisi, and R. Petronzio, \NPB{158}{79}{295};
B. Pendleton and G. Ross, \PLB{98}{81}{291}; C. Hill. \PRD{24}{81}{691};
J. Bagger, S. Dimopoulos, and E. Masso, \NPB{253}{85}{397}.
\bibitem{aspects}S. Kelley, \JL, \DVN, H. Pois, and K. Yuan, \NPB{398}{93}{3}.
\bibitem{DLplb}L. Durand and \JL, \PLB{217}{89}{463}.
\bibitem{GBGS}N. Gray, D. Broadhurst, W. Grafe, and K. Schilcher, Z. Phys.
C{\bf48} (1990) 673.
\bibitem{su5} See \eg, J. Ellis, S. Kelley, and \DVN, \NPB{373}{92}{55}; H.
Arason, \etal, \PRL{67}{91}{2933}; S. Kelley, \JL, and \DVN,
\PLB{274}{92}{387}; V. Barger, M. Berger, and P. Ohman, \PRD{47}{93}{1093}; P.
Langacker and N. Polonsky, \PRD{49}{94}{1454}.
\bibitem{so10} B. Ananthanarayan, G. Lazarides, and Q. Shafi,
\PRD{44}{91}{1613}; S. Kelley, \JL, and \DVN, as in Ref.~\cite{su5}; L. Hall,
R. Ratazzi, and U. Sarid, LBL-33997 (revised March 1994) and SU-ITP-94-15
(hep-ph/9405313).
\bibitem{Kane}C. Kolda, L. Roszkowski, J. Wells, and G. Kane, UM-TH-94-03
(hep-ph/9404253).
\bibitem{Carena}M. Carena, M. Olechowski, S. Pokorski, and C. Wagner,
CERN-TH.7163/94 (hep-ph/9402253).
\bibitem{EWx}L. Ib\'a\~nez and G. Ross, \PLB{110}{82}{215}; K. Inoue, \etal,
Prog. Theor. Phys. 68 (1982) 927; L. Ib\'a\~nez, \NPB{218}{83}{514} and
\PLB{118}{82}{73}; H. P. Nilles, \NPB{217}{83}{366}; J. Ellis, \DVN, and
K. Tamvakis, \PLB{121}{83}{123}; J. Ellis, J. Hagelin, \DVN, and K. Tamvakis,
\PLB{125}{83}{275}; L. Alvarez-Gaum\'e, J. Polchinski, and M. Wise,
\NPB{221}{83}{495}; L. Iba\~n\'ez and C. L\'opez, \PLB{126}{83}{54} and
\NPB{233}{84}{545}; C. Kounnas, A. Lahanas, \DVN, and M. Quir\'os,
\PLB{132}{83}{95} and C. Kounnas, A. Lahanas, \DVN, and M. Quir\'os,
\NPB{236}{84}{438}.
\bibitem{LN}For a review see A. B. Lahanas and D. V. Nanopoulos, Phys. Rept.
{\bf145}~(1987)~1.
\bibitem{FFF} I. Antoniadis, C. Bachas, and C. Kounnas, Nucl. Phys. B
{\bf 289} (1987) 87; I. Antoniadis and C. Bachas, Nucl. Phys. B {\bf298} (1988)
586; H. Kawai, D.C. Lewellen, and S.H.-H. Tye, Phys. Rev. Lett. {\bf57} (1986)
1832; Phys. Rev. D {\bf34} (1986) 3794; Nucl. Phys. B {\bf288} (1987) 1;
R. Bluhm, L. Dolan, and P. Goddard, Nucl. Phys. B {\bf309} (1988) 330;
H. Dreiner, J. L. Lopez, D. V. Nanopoulos, and D. Reiss, Nucl. Phys. B
{\bf 320} (1989) 401.
\bibitem{revamped} I. Antoniadis, J. Ellis, J. Hagelin, and \DVN,
\PLB{231}{89}{65}.
\bibitem{decisive} \JL\ and \DVN, \PLB{251}{90}{73} and \PLB{268}{91}{359}.
\bibitem{tpaper}\JL, \DVN, and \AZ, \PLB{327}{94}{279}.
\end{thebibliography}
\end{document}